\documentclass[12pt]{article}
\usepackage{amsfonts}
\usepackage{amssymb}
\usepackage{amsmath}
\usepackage{theorem}

\newtheorem{theo}{Theorem}[section]
{\theorembodyfont{\rmfamily}
\newtheorem{defin}[theo]{Definition}
}

\newenvironment{proof}{\textit{Proof.}}{\hfill$\square$}

{\theorembodyfont{\rmfamily}

}
{\theorembodyfont{\rmfamily}

}
{\theorembodyfont{\rmfamily}
\newtheorem{remark}[theo]{Remark}
}

\begin{document}

\title{New results on egalitarian values for games with a priori unions%
}
\author{J.C. Gon\c{c}alves-Dosantos$^1$, J.M. Alonso-Meijide$^2$}
\date{\empty}
\maketitle

\begin{abstract}
Several extensions of the equal division value and the equal surplus division value to the family of games with a priori unions are proposed in Alonso-Meijide et al. (2020) ``On egalitarian values for cooperative games with a priori unions.'' TOP 28: 672-688. In this paper we provide new axiomatic characterizations of these values. Furthermore, using the equal surplus division value in two steps, we propose a new coalitional value. The balanced contributions and quotient game properties give rise to a different modification of the equal surplus division value.
\end{abstract}
	
	\footnotetext[1]{Grupo MODES, CITIC and Departamento de Matem\'aticas, Universidade da Coru\~{n}a, Campus de Elvi\~{n}a, 15071 A Coru\~{n}a, Spain. Corresponding author juan.carlos.goncalves@udc.es.}
	\footnotetext[2]{Grupo MODESTYA, Departamento de Estat\'{\i}stica, An\'alise Matem\'atica e
	Optimizaci\'on, Universidade de Santiago de Compostela, Facultade de
	Ciencias, Campus de Lugo, 27002 Lugo, Spain.}



\section{Introduction}

One of the main subjects of study in cooperative game theory is how to divide an existing amount among a set of agents. The Shapley value (Shapley, 1953) is probably the most successful answer to this problem. In this paper, we analyze two different alternatives to the Shapley value: the egalitarian solution (in which the worth of the grand coalition is divided equally among the players), and the equal surplus division value (Driessen and Funaki, 1991) that first allocates individual payoffs to each agent and, after that, the remaining amount is divided equally among them. These values satisfy several good properties and then, van den Brink and Funaki (2009), Chun and Park (2012), van den Brink et al. (2016), Ferri\`eres (2017) and B\`eal et al. (2019), among others, provide many axiomatic characterizations.

Players with similar interests are more likely to act together than others, giving rise to games where cooperation is restricted by an a priori system of unions. For this kind of games, different coalitional values have been analyzed. The first one, proposed by Aumann and Dr$\grave{e}$ze (1974), considers that every player receives the Shapley value of the game played within his union. Owen (1977) defined a different coalitional value (the Owen value) based in the following process. First, the unions play a game among themselves (quotient game), and each one receives a payoff which is shared among its players in a second (internal) game. In both cases, the Shapley value is used to compute the corresponding payoffs. The Owen value coincides with the Shapley value when all unions are singletons, that is, it is a coalitional Shapley value. The Owen value satisfies several good properties as the quotient game and the balanced contribution properties. The quotient game property says that the players of an a priori union receive the amount that this union gets in the quotient game. The balanced contributions property compares the amount obtained by two players of the same union in the original situation with that where one of them leaves this union. In V\'azquez-Brage et al. (1997) the Owen value is characterized as the unique coalitional Shapley value that satisfies the quotient game and balanced contributions properties.

Other values are extended to cooperative games with a priori unions. In the context that concerns us, Alonso-Meijide et al. (2020) extend and characterize the equal division value and the equal surplus division value. For the second value, three alternative ways to adapt it to a priori unions are proposed.

Our aim here is to provide a new axiomatic characterization for the values proposed in Alonso-Meijide et al. (2020) that is able to be compared with the characterization of the Owen value proposed by V\'azquez-Brage et al. (1997). However, none of the extensions of the equal surplus division value satisfies both properties (the quotient game and balanced contributions). Therefore, we propose variants of the properties that allow us to characterize the extensions proposed in Alonso-Meijide et al. (2020).

Further, we obtain a new modification of the equal surplus division value following the Owen process using the equal surplus division value (instead of the Shapley value) in the two steps. Finally, we provide the expression of the coalitional extension of the equal surplus division value satisfying the same properties to those used by Vázquez-Brage et al. (1997) to characterize the Owen value.

\section{Preliminaries}
\label{seccion ed}

\subsection{TU-games and values}
A transferable utility cooperative game (from now on a {TU-game}) is a pair $%
(N,v)$ where $N$ is a finite set of $n$ players, and $v$ is a map from $2^N$
to $\mathbb{R}$ with $v(\emptyset)=0$, that is called the characteristic
function of the game. In the sequel, $\mathcal{G}_N$ will denote the family
of all TU-games with player set $N$ and $\mathcal{G}$ the family of all
TU-games. A {value} for TU-games is a map $f$ that assigns to every TU-game $%
(N,v)\in\mathcal{G}$ a vector $f(N,v)=(f_i(N,v))_{i\in N}\in\mathbb{R}^N$.

Well-known values for TU-games are the egalitarian values. The {\ equal division value} $%
ED $ distributes $v(N)$ equally among the players in $N$. Formally, the
equal division value $ED$ is defined for every $(N,v)\in\mathcal{G}$, and every $i\in N$ by 
\begin{equation*}
ED_{i}(N,v)=\frac{v(N)}{n}.
\end{equation*}

\bigskip The equal surplus division value $ESD$ is defined
for every $(N,v)\in \mathcal{G}$, and every $i\in N$ by 
\begin{equation*}
ESD_{i}(N,v)=v(i)+\frac{v^{0}(N)}{n}
\end{equation*}%
where $v^{0}(N)=v(N)-\sum_{i\in N}v(i)$. Notice that $ESD$ is a variant of $%
ED$ in which we first allocate $v(i)$ to each player $i$, and then
distribute $v^{0}(N)$ among the players using $ED$. $ESD$ is a reasonable
alternative to $ED$ for situations where individual benefits and joint
benefits are neatly separable.

Alternative values for TU-games are the Shapley (Shapley 1953) and Banzhaf value (Banzhaf 1964).

\subsection{Games with a priori unions}
We denote by $P(N)$ the set of all partitions of $N$. Then, a {TU-game with a
priori unions} is a triplet $(N,v,P)$ where $(N,v)\in\mathcal{G}$, $%
P=\{P_1,\dots,P_m\}\in P(N)$ and $P_k\in P$ is called a priori union for all $k\in M$ with $M=\{1,...,m\}$. The set of TU-games with a priori unions and
with player set $N$ will be denoted by $\mathcal{G}_N^{U}$, and the set of
all TU-games with a priori unions by $\mathcal{G}^{U}$. A {value for
TU-games with a priori unions} is a map $g$ that assigns to every $(N,v,P)\in%
\mathcal{G}^U$ a vector $g(N,v,P)=(g_i(N,v,P))_{i\in N}\in\mathbb{R}^N$. 

Two examples of values for TU-games with a priori unions are the Owen value (Owen 1977), and the Banzhaf-Owen value (Owen 1981).

Given $(N,v,P)\in \mathcal{G}^U$ with $P=\{P_1,\dots,P_m\}\in P(N)$, the quotient game of $(N,v,P)$ is the TU-game $(M,v/P)$ where
\[(v/P)(R)=v\left(\cup_{r\in R}P_r\right) \text{ for all } R\subseteq M.\]

We say that a value for TU-games with a priori unions $g$ is a \emph{%
	coalitional equal division value (CED)} if, for any TU-game $(N,v)\in\mathcal{G}$,
it holds that 
\begin{equation*}
g(N,v,P^n)=ED(N,v),
\end{equation*}
where

$$P^n=\{\{ 1\},\dots,\{ n\}\}.$$ 
 
Using similar concepts, the Owen value is a
coalitional Shapley value, and Banzhaf-Owen value is a coalitional Banzhaf value.

\subsection{Coalitional values in two steps}
\label{owenmethod}

Given $\left( N,v,P\right)\in\mathcal{G}_N^U$ with $P=\{P_{1},\dots ,P_{m}\}$ and a coalition $S\subseteq P_{r}$, the modified game of $\left( N,v,P\right)$  is defined as $(M,u_{r,S})$ where

\begin{equation}
u_{r,S}\left( H\right) =\left\{ 
\begin{array}{cc}
v\left( \cup _{k\in H}P_{k}\right) & \text{if }r\notin H \\ 
v\left( \cup _{k\in H\backslash r}P_{k}\cup S\right) & \text{if }r\in H%
\end{array}%
\right.
\label{eq1}
\end{equation}
for all $H\subseteq M.$\ That is, the modified game $(M,u_{r,S})$, is defined based on the game $(N,v,P)$ where each player $k$ with $k\neq r$ is the union $P_{k}$, and the player  $r$ is the coalition $S$.

Using the modified game $(M,u_{r,S})$ and a value $f$ for TU-games, the reduced game $\left(P_r,w_{r}\right)$ is a TU-game with set of players $P_{r}$ and characteristic function

\begin{equation}
w_{r}\left( S\right) =f_{r}\left( M,u_{r,S}\right)
\label{eq2}
\end{equation}
for any $S\subseteq P_{r}$.

Finally, a value  $g$ for TU-games with a priori unions is obtained
reapplying the value $f$ over $(P_r,w_{r})$. That is
\begin{equation}
g_{i}\left( N,v,P\right) =f_{i}\left( P_{r},w_{r}\right)
\label{eq3}
\end{equation}
for all $i\in P_{r}$.

We call Owen procedure to that described in (\ref{eq2}) and (\ref{eq3}) to obtain a coalitional value g for TU games with a priori unions using a value f for TU games.

The Owen value is the result of applying the Owen procedure using the Shapley value (Owen 1977), and the Banzhaf-Owen value is the result of applying the Owen procedure using the Banzhaf value (Owen 1981).

A similar approach to obtain coalitional values in two steps is presented in G\'omez-R\'ua and Vidal-Puga (2010). They propose different coalitional values, in one of them, the pay-offs obtained by the unions are given using a weighted Shapley value, with weights given by the size of unions. 

\section{The equal division value for TU-games with a priori unions}
\label{edu}


Alonso-Meijide et al. (2020) define the equal division value for TU-games with a priori unions as the natural extension of the equal division value to the set $\mathcal{G}^{U}$.

\begin{defin} (Alonso-Meijide et al. 2020)
The equal division value for TU-games with a priori unions $ED^U$ is defined
by 
\begin{equation*}
ED^U_{i}\left(N,v,P\right)=\frac{v(N)}{mp_k}
\end{equation*}
for all $(N,v,P)\in\mathcal{G}^U$ with $P=\{P_1,\dots,P_m\}$
and $i\in P_k$ where $p_k$ denotes the cardinal of $P_k$.
\end{defin}

Alonso-Meijide et al. (2020) characterize the equal division value for TU-games with a priori unions using, among others, two properties of symmetry and additivity. In this section we provide a second axiomatic characterization of the equal division value for TU-games with a priori unions using the concept of coalitional equal division value and two additional properties: the quotient game property (Winter, 1992) and the balanced contributions in the unions property (V\'azquez-Brage et al., 1996). First, we remember these two properties.


\bigskip \noindent \textbf{Quotient Game Property (QGP).} A value $g$ for
TU-games with a priori unions satisfies the quotient game property if, for
all $( N,v,P)\in\mathcal{G}_N^U$ with $P=\{P_1,\dots,P_m\}$, it holds that 
\begin{equation*}
\sum_{i\in P_k}g_{i}\left( N,v,P\right) =g_{k}\left( M,v/P,P^{m}\right)
\end{equation*}
for all $P_k\in P$, where
 $(M,v/P)$ is the quotient game of $(N,v,P)$.

\bigskip \noindent\textbf{Balanced Contributions in the Unions (BCU).} A
value $g$ for TU-games with a priori unions satisfies balanced contributions
in the unions if, for all $( N,v,P)\in\mathcal{G}_N^U$ and all $i,j\in P_k$
with $P_k\in P$, it holds that 
\begin{equation*}
g_{i}\left( N,v,P\right)-g_{i}\left( N,v,P_{-j}\right) =g_{j}\left(
N,v,P\right)-g_{j}\left( N,v,P_{-i}\right)
\end{equation*}
where $P_{-l}$ denotes the partition $\{P_1,\dots
,P_{k-1},P_k\setminus\{l\},\{l\},P_{k+1},\dots ,P_m \}$ for all $l\in P_k$.

\bigskip

If a value satisfies the quotient game property then the total amount received by the players of an union coincides with the amount that this union obtains in the game played by the unions (the quotient game). For example, the Owen value satisfies this property but the Banzhaf-Owen value does not. The balanced contributions in the unions property compares the payoff obtained by a player $i$ in the original game $(N,v,P)$ and in the game when a player $j$ of the same union decides to leave the union and stay alone $(N,v,P_{-j})$. This property establishes that the difference between the payoffs obtained by player $i$ in the two previous games, coincides with the same difference for player $j$ when player $i$ leaves the union. This property is a particular case of the splitting property (Casajus, 2009). In the case of the splitting property this difference is the same considering any game $(N,v,P')$ where the partition $P'$ is finer than $P$.

V\'azquez-Brage et al. (1997) prove that the Owen value is the unique coalitional Shapley value that satisfies the properties of quotient game and balanced contributions in the unions. In a similar way, Alonso-Meijide and Fiestras-Janeiro (2002) characterize the coalitional Banzhaf value as the unique coalitional Banzhaf value that satisfies the properties of quotient game and balanced contributions in the unions. In the same spirit, we present a characterization of the equal division value for TU-games with a priori unions.

The mathematical arguments of some of the proofs presented in this paper are similar to the proofs of previous papers, and they are relegated to the Appendix. They share the ideas of quotient game and balanced contributions properties joint to the concept of coalitional value to show the unicity of the solutions.

\begin{theo}
$ED^U$ is the unique coalitional equal division value satisfying QGP and BCU.
\label{th3.2}
\end{theo}

\bigskip

In the previous theorem ``Coalitional equal division value'' can be stated as a third property. Moreover, ``Coalitional equal division value'' could be replaced by any set of properties that characterizes the equal division value in the family of TU-games by adding the mention of the trivial coalition structure.

The $ED^U$ value is an extension of $ED$ for TU games with a priori unions quite intuitive and natural. Moreover, let us check that it is the value obtained by the procedure to obtain coalitional values in two steps proposed in Owen (1977) described in Subsection \ref{owenmethod}.

\begin{theo}
	The equal division value with a priori unions $ED^U$ is the result of applying the Owen procedure using the equal division value $ED$. 
	\label{edow}
\end{theo}

\section{Three equal surplus division values for TU-games with a priori unions}
\label{seccion esd}

In Alonso-Meijide et al. (2020), three alternative ways for extending the equal surplus division value to TU-games with a priori unions are proposed. In this section we provide new characterizations of these coalitional values. 

\subsection{The equal surplus division value 1}

The equal surplus division value $1$ divides the
value of the grand coalition in the quotient game using the equal surplus
division value and then divides the amount assigned to each union equally
among its members.

\begin{defin} (Alonso-Meijide et al. 2020)
The equal surplus division value (one) for TU-games with a priori unions $%
ESD1^U$ is defined by 
\begin{equation*}
ESD1^U_{i}\left(N,v,P\right)=\frac{(v/P)(k)}{p_k}+\frac{(v/P)^0(M)}{mp_k}=%
\frac{v(P_k)}{p_k}+\frac{v(N)-\sum_{l\in M}v(P_l)}{mp_k}
\end{equation*}
for all $(N,v,P)\in\mathcal{G}^U$ with $P=\{P_1,\dots,P_m\}$
and $i\in P_k$.
\end{defin}

\bigskip

Notice that it is easy to check that $ESD1^U$ is a
\emph{coalitional equal surplus division value (CESD)}, in the sense that 
\begin{equation*}
ESD1^U(N,v,P^n)=ESD(N,v)
\end{equation*}
for all $(N,v)\in\mathcal{G}$. 

We use this feature to provide a new characterization of $ESD1^U$ in the remainder of this subsection.  First let us define a new property.

\bigskip

\noindent \textbf{Equality Inside Unions (EIU).} A value $g$ for TU-games
with a priori unions satisfies equality inside unions if, for all $%
(N,v,P)\in\mathcal{G}^U$, all $P_k\in P$, and all $i,j\in P_k$, it holds
that $g_i(N,v,P)-g_j(N,v,P)=0$.

\bigskip

The previous property takes up the idea of egalitarianism inside unions. This property considers that players within a union are willing to show strict equality. It could be seen as a stronger version of the symmetry property, in the sense that all players belong to the same union obtain the same pay-off, without taking into account the characteristic function of the game. 

\begin{theo}
	$ESD1^U$ is the unique coalitional equal surplus division value for TU-games
	with a priori unions satisfying QGP and EIU. 
	\label{th6}
\end{theo}

\bigskip

It is very easy to prove that the equal division value for TU-games with a priori unions satisfies EIU. Moreover, EIU could replace BCU in Theorem \ref{th3.2} to obtain a new characterization of the equal division value for TU-games with a priori unions.

\subsection{The equal surplus division value 2}

The equal surplus division value $2$ divides again the value of the grand coalition in the
quotient game using the equal surplus division value; then it distributes
the amount $v(P_k)$ assigned to each union $P_k$ giving $v(i)$ to each
player $i\in P_k$ and dividing $v(P_k)-\sum_{j\in P_k}v(j)$ equally among
the players in $P_k$.

\begin{defin} (Alonso-Meijide et al. 2020)
The equal surplus division value (two) for TU-games with a priori unions $%
ESD2^U$ is defined by 
\begin{equation*}
ESD2^U_{i}\left(N,v,P\right)=v(i)+\frac{v(P_k)-\sum_{j\in P_k}v(j)}{p_k}+%
\frac{v(N)-\sum_{l\in M}v(P_l)}{mp_k}
\end{equation*}
for all $(N,v,P)\in\mathcal{G}^U$ with $P=\{P_1,\dots,P_m\}$ and $i\in P_k$.
\end{defin}

\bigskip

Notice that it is easy to check that $ESD2^U$ is a
\emph{coalitional equal surplus division value (CESD)}, in the sense that 
\begin{equation*}
ESD2^U(N,v,P^n)=ESD(N,v)
\end{equation*}
for all $(N,v)\in\mathcal{G}$. 

We use this feature to provide a  new characterization of $ESD2^U$ in the remainder of this subsection.  First let us define a new property.

\bigskip

\noindent \textbf{Difference Maintenance of Individual Values Inside Unions
	(DMIVIU).} A value $g$ for TU-games with a priori unions satisfies
difference maintenance of individual values inside unions if, for all $%
(N,v,P)\in\mathcal{G}^U$, all $P_k\in P$, and all $i,j\in P_k$, it holds
that $g_i(N,v,P)-g_j(N,v,P)=v(i)-v(j)$.

\bigskip

This property is similar to EIU, in the sense that DMIVIU is a stronger version of the symmetry property, but not so stronger that EIU. In this case, the difference between the pay-offs of two players of the same union coincides with the difference between the amounts given by the characteristic function of the game to individual coalitions, without taking account the amounts given to coalitions with two or more players. It is immediate that in the case of zero normalized games DMIVIU is equivalent to EIU.

\begin{theo}
	$ESD2^U$ is the unique coalitional equal surplus division value for TU-games
	with a priori unions satisfying QGP and DMIVIU. \label{th7}
\end{theo}

\subsection{The equal surplus division value 3}

Finally, the equal surplus division value $3$ assigns $v(i)$ to each player $i$ and then
divides $v^0(N)$ among the players using $ED^U$.

\begin{defin} (Alonso-Meijide et al. 2020)
The equal surplus division value (three) for TU-games with a priori unions $%
ESD3^U$ is defined by 
\begin{equation*}
ESD3^U_{i}\left(N,v,P\right)=v(i)+ED^U(N,v^0,P)=v(i)+ \frac{v(N)-\sum_{j\in
N}v(j)}{mp_k}
\end{equation*}
for all $(N,v,P)\in\mathcal{G}^U$ with $P=\{P_1,\dots,P_m\}$
and $i\in P_k$.
\end{defin}


\bigskip

Notice that it is easy to check that $ESD3^U$ is a
\emph{coalitional equal surplus division value (CESD)}, in the sense that 
\begin{equation*}
ESD3^U(N,v,P^n)=ESD(N,v)
\end{equation*}
for all $(N,v)\in\mathcal{G}$. 

We use this feature to provide a new characterization of $ESD3^U$ in the remainder of this section. Nevertheless $ESD3^U$ does not satisfy QGP, so in order to do it we
introduce a new modified of quotient game.

Take $(N,v,P)\in \mathcal{G}^U$ with $P=\{P_1,...,P_m\}$ and denote $%
M=\{1,...,m\}$. The quotient* game of $(N,v,P)$ is the TU-game $(M,\bar{v}/P)$ where 
\begin{equation*}
(\bar{v}/P)(R) =\left\{ 
\begin{array}{cc}
\displaystyle \sum_{k\in R}\sum_{i\in P_k}v(i) & \text{if } R\subset M \\ 
v(N) & \text{if }R=M%
\end{array}%
\right.
\end{equation*}

\bigskip \noindent \textbf{Quotient* Game Property (Q*GP).} A value $g$
for TU-games with a priori unions satisfies the quotient* game property if,
for all $(N,v,P)\in \mathcal{G}^{U}$ with $P=\{P_{1},...,P_{m}\}$, it holds that 
\begin{equation*}
\sum_{i\in P_{k}}g_{i}\left( N,v,P\right) =g_{k}\left( M,\bar{v}/P,P^{m}\right)
\end{equation*}%
for all $P_k\in P$, where $(M,\bar{v}/P)$ is the quotient* game of $(N,v,P)$.
\bigskip

\begin{theo}
$ESD3^U$ is the unique coalitional equal surplus division value for TU-games
with a priori unions satisfying Q*GP and BCU. \label{th9}
\end{theo}

\bigskip

\section{Two new extensions of the Equal surplus division value}
\label{newvalues}

In this section, we introduce two new extensions of the equal surplus division value for TU-games with a priori unions. One of them is the value obtained applying the Owen procedure using the equal surplus division value. The other value is an extension of the equal surplus division value that satisfies the quotient game and balanced contributions in the unions properties.

\subsection{Coalitional value using Equal surplus division value in two steps}

The first new extension is obtained applying the procedure in two steps using the equal surplus division value. The first part of the value coincides with the $ESD2^U$ as we will mention later. In the second part, it allocates the difference between the average value of the players in the union and the value of the player, then the difference between the value of the player's contribution to the grand coalition minus the union and the average contribution of the players in the union to the grand coalition minus the union, all divided by the total of the unions.

\begin{defin} 
	The equal surplus division value (four) for TU-games with a priori unions $%
	ESD4^U$ is defined by 
	\begin{align*}
	&ESD4_{i}^{U}\left( N,v,P\right) =v(i)+\frac{v(P_k)-\sum_{j\in P_k}v(j)}{p_k}+%
	\frac{v(N)-\sum_{l\in M}v(P_l)}{mp_k} +\\ &\frac{1}{m}\left( \sum_{t\in P_{k}}\frac{v\left( t\right) 
	}{p_{k}}-v\left( i\right) \right)+
	\frac{1}{m}\left( v\left( \cup _{r\in M\backslash k}P_{r}\cup i\right)
	-\sum_{t\in P_{k}}\frac{v\left( \cup _{r\in M\backslash k}P_{r}\cup t\right) 
	}{p_{k}}\right)\\
	\end{align*}
	for all $(N,v,P)\in\mathcal{G}^U$ with $P=\{P_1,\dots,P_m\}$
	and  $i\in P_k$.
\end{defin}

Using again the procedure proposed by Owen (1977) shown in Subsection \ref{owenmethod} where now $ESD$ is used in the games (\ref{eq2}) and (\ref{eq3}), we can obtain the solution $g=ESD4^{U}$. Let us see this in the next theorem.

\bigskip

\begin{theo}
	The equal surplus division value with a priori unions $ESD4^U$ is the result of applying the Owen procedure using the equal surplus division value $ESD$.
	\label{esdow}
\end{theo}

\bigskip 

\begin{remark}
Note that $ESD4$ can be written in terms of $ESD2$
\begin{align*}
&ESD4_{i}^{U}\left( N,v,P\right) =ESD2^U_{i}\left( N,v,P\right) + \frac{1}{m}\left( \sum_{t\in P_{r}}\frac{v\left( t\right) 
}{p_{r}}-v\left( i\right) \right)+\\
&\frac{1}{m}\left( v\left( \cup _{k\in M\backslash r}P_{k}\cup i\right)
-\sum_{t\in P_{r}}\frac{v\left( \cup _{k\in M\backslash r}P_{k}\cup t\right) 
}{p_{r}}\right)\\
\end{align*}
for all $(N,v,P)\in\mathcal{G}^U$ with $P=\{P_1,\dots,P_m\}$
and  $i\in P_k$.
\end{remark}

\bigskip 

It is easy to check that
\begin{equation*}
ESD4_{i}^{U}\left( N,v,P^{n}\right) =ESD_{i}(N,v)
\end{equation*}
 and
\begin{equation*}
\sum_{i\in P_{r}}ESD4_{i}^{U}\left( N,v,P\right) =ESD_{r}(M,v/P),
\end{equation*}%
that is $ESD4^U$  is a coalitional equal surplus division value and it satisfies the quotient game property.

The value $ESD4$ is a coalitional equal surplus division value that satisfies quotient game. To characterize $ESD4^U$ in the proposed context, let us define the following property.

\bigskip 

\textbf{Balanced contributions due to the players abandonment in the union (BCPA).} A value $g$ for TU-games with a priori unions satisfies BCPA if, for all $(N,v,P)\in \mathcal{G}^{U}$ and all $i,j\in P_{k}$ with $%
P_{k}\in P$, it holds that 
\begin{equation*}
g_{i}\left( N,v,P\right) -g_{i}\left( N\backslash P_{k}\cup i,v_{N\backslash
P_{k}\cup i},P\backslash P_{k}\cup \{i\}\right) =
\end{equation*}%
\begin{equation*}
g_{j}\left( N,v,P\right) -g_{j}\left( N\backslash P_{k}\cup j,v_{N\backslash
P_{k}\cup j},P\backslash P_{k}\cup \{j\}\right) 
\end{equation*}
where the game $\left( N\backslash P_{k}\cup i,v_{N\backslash P_{k}\cup
	i},P\backslash P_{k}\cup \{i\}\right) $ is defined as $v_{N\backslash P_{k}\cup
	i}(S)=v(S)$ for all $S\subseteq N\backslash P_{k}\cup i$.

\bigskip 

This new property says that given two players in the same union, they get the same difference between the pay-off of the original game and the pay-off of the game where all the players of the union leave. This property has a similar interpretation to the property of balanced contributions.

\begin{theo}
	$ESD4^U$ is the unique coalitional equal surplus division value for TU-games
	with a priori unions satisfying QGP and BCPA. \label{th10}
\end{theo}

\bigskip

\subsection{Coalitional equal surplus division value satisfying balanced contributions and quotient game}

In this section we define a value for TU-games with a priori unions that extends the equal surplus division value and satisfies the quotient game property and balanced contributions in the unions.  The first part of the value coincides with the $ESD1^U$ as we will mention later; and a weighted difference between the value of the subsets of the union that contain the player minus the subsets which do not.

\begin{defin} 
	The equal surplus division value (five) for TU-games with a priori unions $%
	ESD5^U$ is defined by 
\begin{equation*}
ESD5_{i}^{U}\left( N,v,P\right) =\frac{v(P_{k})}{p_{k}}+\frac{v\left(
	N\right) -\sum_{l\in M}v\left( P_{l}\right) }{mp_{k}}+\sum_{\substack{ %
		T\subset P_{k} \\ i\in T}}\frac{P^{m,p_{k},t}}{t}v\left(
T\right) -\sum_{\substack{ T\subset P_{k} \\ i\notin T}}\frac{P^{m,p_{k},t}}{p_k-t}v\left( T\right) 
\end{equation*}
for all $(N,v,P)\in\mathcal{G}^U$ with $P=\{P_1,\dots,P_m\}$
and $i\in P_k$; where $t=|T|$ and
\begin{align*}
&P^{m,p_{k},t}=\frac{1}{2}& \text{ if } p_k=2\text{ and } t=1,\\	
&P^{m,p_{k},t}=\frac{1}{p_k}\left(1+\sum_{j=1}^{p_k-2}\frac{1}{m+j}\right)& \text{ if } p_k>2\text{ and } t=1,\\
&P^{m,p_{k},t}=\frac{m}{(m+1)p_k}& \text{ if } p_k>2\text{ and } t=p_k-1,\\
&P^{m,p_{k},t}=\frac{m+(z-1)}{(p_k-(z-1))(m+z)}\left(\sum_{j=0}^{z-2}\frac{p_k-j-t}{p_k-j}\right)& \\
&\text{ if } p_k>3\text{ and } t=(p_k-z)\text{ such that }z\in\{2,...,p_k-2\}.\\	
\end{align*}
\end{defin}

\begin{remark}
	Note that $ESD5$ can be written in terms of $ESD1$
	\begin{align*}
	&ESD5_{i}^{U}\left( N,v,P\right) =ESD1^U_{i}\left( N,v,P\right) +\sum_{\substack{ %
			T\subset P_{k} \\ i\in T}}\frac{P^{m,p_{k},t}}{t}v\left(
	T\right) -\sum_{\substack{ T\subset P_{k} \\ i\notin T}}\frac{P^{m,p_{k},t}}{p_k-t}v\left( T\right).\\
	\end{align*}
	for all $(N,v,P)\in\mathcal{G}^U$ with $P=\{P_1,\dots,P_m\}$
	and  $i\in P_k$.
\end{remark}

In the last result of this work, we characterized the $ESD5$ value in the same spirit as an axiomatic characterization of the Owen value given in V\'azquez-Brage et al. (1997).

\begin{theo}
	\label{th12}
$ESD5^U$ is the unique coalitional equal surplus division value satisfying QGP and BCU.

\end{theo}

\bigskip

Table \ref{tab:1} shows in a summarised form the properties fulfilled by all the values. Note that all theorems are based on independent properties. The logical demonstration of the independence of the axioms is provided as an online supplement to readers who wish to request them.

\begin{table}[h!]
	\begin{center} 
	\begin{tabular}{ c | ccccccccc}
		&$CED$&$CESD$&$QGP$&$Q^*GP$&$BCU$&$EIU$&$DMIVIU$&$BCPA$\\ \hline
		$ED^U$ &$\checkmark$&$-$&$\checkmark$& $-$&$\checkmark$&$-$&$-$&$-$  \\
		$ESD1^U$&$-$&$\checkmark$&$\checkmark$& $-$&$-$&$\checkmark$&$-$&$-$\\
		$ESD2^U$&$-$&$\checkmark$&$\checkmark$& $-$&$-$&$-$&$\checkmark$&$-$\\
		$ESD3^U$&$-$&$\checkmark$&$-$& $\checkmark$&$\checkmark$&$-$&$-$&$-$\\
		$ESD4^U$&$-$&$\checkmark$&$\checkmark$& $-$&$-$&$-$&$-$&$\checkmark$\\
		$ESD5^U$&$-$&$\checkmark$&$\checkmark$& $-$&$\checkmark$&$-$&$-$&$-$\\
	\end{tabular}
	\end{center}
\caption{Properties satisfied by all the values}
\label{tab:1}  
\end{table}

\section*{Acknowledgements}

\noindent This work has been supported by the ERDF, the MINECO/AEI grants
MTM2017-87197-C3-1-P, MTM2017-87197-C3-3-P, and by the Xunta de Galicia
(Grupos de Referencia Competitiva ED431C-2016-015 and ED431C-2017/38). CITIC as Centro de Investigaci\'on do Sistema universitario de Galicia is financed by  Conseller\'ia de Educaci\'on,  Universidade e Formaci\'on Profesional of Xunta de Galicia through the Fondo Europeo de Desenvolvemento Rexional (FEDER) with 80\%, Programa operativo FEDER Galicia 2014-2020 and the remaining 20\% by the Secretar\'ia  Xeral de Universidades (Ref. ED431G 2019/01)

\section*{Appendix}

Here, the reader can find the proofs of the theorems stated in this paper.

\bigskip

Proof of Theorem \ref{th3.2}.

\bigskip

\begin{proof}
	Take a TU-game with a priori unions $(N,v,P)\in \mathcal{G}^U$ such that $P=\{P_1,...,P_m\}$ and denote $M=\{1,...,m\}$. Let us check that $ED^{U}$ satisfies QGP. For all $k\in M$, we have that 
	\begin{equation*}
	\sum_{i\in P_{k}}ED_{i}^{U}\left( N,v,P\right) =\sum_{i\in P_{k}}\frac{v(N)}{%
		mp_{k}}=\frac{v(N)}{m}
	\end{equation*}%
	and 
	\begin{equation*}
	ED_{k}^{U}\left( M,v/P,P^{m}\right) =\frac{(v/P)(M)}{m}=\frac{v(N)}{m}.
	\end{equation*}%
	
	Let us check that $ED^{U}$ satisfies BCU. For all $i,j\in P_{k}$, we have
	that 
	\begin{equation*}
	ED_{i}^{U}\left( N,v,P\right) -ED_{i}^{U}\left( N,v,P_{-j}\right) =\frac{v(N)%
	}{mp_{k}}-\frac{v(N)}{(m+1)(p_{k}-1)}
	\end{equation*}%
	and 
	\begin{equation*}
	ED_{j}^{U}\left( N,v,P\right) -ED_{j}^{U}\left( N,v,P_{-i}\right) =\frac{v(N)%
	}{mp_{k}}-\frac{v(N)}{(m+1)(p_{k}-1)}.
	\end{equation*}%
	
	Finally, the uniqueness is proven in an analogous way as the uniqueness in
	Theorem 2 of V\'{a}zquez-Brage et al. (1997). Let us suppose that there exist two different coalitional equal division values $f^{1}$ and $f^{2}$ satisfying Quotient Game property and Balanced Contributions in the Unions. We can find a coalitional game $(N,v,P)$, where $P$ is a maximal number of unions such that $f^{1}(N,v,P)\neq f^{2}(N,v,P)$. Taking into account that $f^{1}$ and $f^{2}$ satisfying Quotient Game property, $\forall P_k\in P$ and $l\in\{1,2\}$, we have
	\[\sum_{i\in P_k}f^{l}_{i}\left( N,v,P\right) =f^{l}_{k}\left( M,v/P,P^{m}\right).\]
	But $f^{1}$ and $f^{2}$ are coalitional equal division values, then
	\begin{equation}
	\sum_{i\in P_k}f^{1}_{i}\left( N,v,P\right) =\sum_{i\in P_k}f^{2}_{i}\left( N,v,P\right)=ED_{k} \left( M,v/P\right).
	\label{qg}
	\end{equation}
	
	If $P_k$ is such that $|P_k|=1$, i.e. $P_k=\{i\}$, then
	\[f^{1}_{i}\left( N,v,P\right) =f^{2}_{i}\left( N,v,P\right)\]
	However, if $|P_k|>1$, for any $i,j\in P_k$, we have by Balanced Contributions in the Unions
	\[f^{l}_{i}\left( N,v,P\right) -f^{l}_{j}\left( N,v,P\right)=f^{l}_{i}\left( N,v,P_{-j}\right)-f^{l}_{j}\left( N,v,P_{-i}\right)\]
	for all $l\in\{1,2\}$. Therefore the maximality of $P$ implies that
	\[f^{1}_{i}\left( N,v,P\right) -f^{1}_{j}\left( N,v,P\right)=f^{2}_{i}\left( N,v,P\right) -f^{2}_{j}\left( N,v,P\right)\]
	and we have
	\[f^{1}_{i}\left( N,v,P\right) -f^{2}_{i}\left( N,v,P\right)=A^{k}\]
	for all $i\in P_k$. By $\ref{qg}$ we have to $A^{k}=0$ and
	\[f^{1}_{i}\left( N,v,P\right) =f^{2}_{i}\left( N,v,P\right)\]
	Thus $f^{1}\left( N,v,P\right) =f^{2}\left( N,v,P\right)$, and we have proven the uniqueness.
\end{proof}

\bigskip

\bigskip

Proof of Theorem \ref{edow}.

\bigskip

\begin{proof}
	Take a TU-game with a priori unions $(N,v,P)\in \mathcal{G}^U$ such that $P=\{P_1,...,P_m\}$ and denote $M=\{1,...,m\}$. Given a coalition $S\subseteq P_r$, we can obtained the reduced game (\ref{eq2}) applying $ED$ to the modified game (\ref{eq1}),
	\begin{equation*}
	w_{r}\left( S\right) =ED_{r}\left( M,u_{r,S}\right) =\frac{u_{r,S}\left(
		M\right) }{m}=\frac{v\left( \cup _{k\in H\backslash r}P_{k}\cup S\right) }{m}
	\end{equation*}
	
	Again, if we reapply $ED$ to the reduced game (\ref{eq2}) as (\ref{eq3}), for all player $i\in P_{r}$ we obtain
	\begin{equation*}
	ED_{i}\left( P_{r},w_{r}\right) =\frac{w_{r}\left( P_{r}\right) }{p_{r}}=%
	\frac{v\left( \cup _{k\in H\backslash r}P_{k}\cup P_{r}\right) /m}{p_{r}}=%
	\frac{v\left( N\right) }{mp_{r}}=ED_{i}^{U}\left( N,v,P\right).
	\end{equation*}
\end{proof}

\bigskip

\bigskip

Proof of Theorem \ref{th6}.

\bigskip

\begin{proof}
	Take a TU-game with a priori unions $(N,v,P)\in \mathcal{G}^U$ such that $P=\{P_1,...,P_m\}$ and denote $M=\{1,...,m\}$. Let us check that $ESD1^U$ satisfies QGP. For all $k\in M$, we have that 
	
	\begin{align*}
	\sum_{i\in P_k}ESD1^U_{i}\left( N,v,P\right)&=
	\sum_{i\in P_k}\left(\frac{%
		v(P_k)}{p_k}+\frac{v(N)-\sum_{l\in M}v(P_l)}{mp_k}\right)\\
	&=v(P_k)+\frac{%
		v(N)-\sum_{l\in M}v(P_l)}{m}
	\end{align*}
	and 
	\begin{align*}
	ESD1^U_{k}\left( M,v/P,P^{m}\right)&=\frac{(v/P)(k)}{1}+\frac{%
		(v/P)(M)-\sum_{l\in M}(v/P)(l)}{m}\\
	&=v(P_k)+\frac{v(N)-\sum_{l\in M}v(P_l)}{m}.
	\end{align*}
	
	Let us check that $ESD1^U$ satisfies EIU. For all $i,j\in P_k$, we have that 
	\begin{equation*}
	ESD1^U_{i}\left( N,v,P\right)-ESD1^U_{j}\left( N,v,P\right) =0.
	\end{equation*}
	
	Finally, the uniqueness is proven in an analogous way as the uniqueness in
	Theorem 2 of V\'azquez-Brage et al. (1997).
\end{proof}

\bigskip

Proof of Theorem \ref{th7}.

\bigskip

\begin{proof}
	Take a TU-game with a priori unions $(N,v,P)\in \mathcal{G}^U$ such that $P=\{P_1,...,P_m\}$ and denote $M=\{1,...,m\}$. Let us check that $ESD2^U$ satisfies QGP. For all $k\in M$, we have that 
	\begin{align*}
	\sum_{i\in P_k}ESD2^U_{i}\left( N,v,P\right)&=\sum_{i\in P_k}\left(v(i)+%
	\frac{v(P_k)-\sum_{j\in P_k}v(j)}{p_k}+\frac{v(N)-\sum_{l\in M}v(P_l)}{mp_k}%
	\right) \\
	&=\sum_{i\in P_k}v(i)+v(P_k)-\sum_{j\in P_k}v(j)+\frac{v(N)-\sum_{l\in
			M}v(P_l)}{m} \\
	&=v(P_k)+\frac{v(N)-\sum_{l\in M}v(P_l)}{m}
	\end{align*}
	and 
	\begin{align*}
	ESD2^U_{k}\left( M,v/P,P^{m}\right)&=(v/P)(k)+\frac{(v/P)(k)-(v/P)(k)}{1}+%
	\frac{(v/P)(M)-\sum_{l\in M}(v/P)(l)}{m} \\
	&=v(P_k)+\frac{v(N)-\sum_{l\in M}v(P_l)}{m}.
	\end{align*}
	
	Let us check that $ESD2^U$ satisfies DMIVIU. For all $i,j\in P_k$, we have
	that 
	\begin{equation*}
	ESD2^U_{i}\left( N,v,P\right)-ESD2^U_{j}\left( N,v,P\right) =v(i)-v(j).
	\end{equation*}
	
	Finally, the uniqueness is proven in an analogous way as the uniqueness in
	Theorem 2 of V\'azquez-Brage et al. (1997).
\end{proof}

\bigskip

\bigskip

Proof of Theorem \ref{th9}.

\bigskip

\begin{proof}
	Take a TU-game with a priori unions $(N,v,P)\in \mathcal{G}^U$ such that $P=\{P_1,...,P_m\}$ and denote $M=\{1,...,m\}$. Let us check that $ESD3^{U}$ satisfies Q*GP. For all $k\in M$, we have that 
	\begin{align*}
	\sum_{i\in P_{k}}ESD3_{i}^{U}\left( N,v,P\right) & =\sum_{i\in P_{k}}\left(
	v(i)+\frac{v(N)-\sum_{j\in N}v(j)}{mp_{k}}\right) \\
	& =\sum_{i\in P_{k}}v(i)+\frac{v(N)-\sum_{j\in N}v(j)}{m}
	\end{align*}%
	and  
	\begin{align*}
	ESD3_{k}^{U}\left( M,\bar{v}/P,P^{m}\right) & =(\bar{v}/P)(k)+\frac{(\bar{v}/P)(M)-\sum_{l\in M}(\bar{v}/P)(l)}{m%
	} \\
	& =\sum_{i\in P_{k}}v(i)+\frac{v(N)-\sum_{k\in M}\sum_{j\in P_{k}}v(j)}{m} \\
	& =\sum_{i\in P_{k}}v(i)+\frac{v(N)-\sum_{j\in N}v(j)}{m}.
	\end{align*}%
	
	Let us check that $ESD3^{U}$ satisfies BCU. For all $i,j\in P_{k}$, we have
	that
	\begin{equation*}
	ESD3_{i}^{U}\left( N,v,P\right) -ESD3_{i}^{U}\left( N,v,P_{-j}\right) =\frac{%
		v(N)-\sum_{j\in N}v(j)}{mp_{k}}-\frac{v(N)-\sum_{j\in N}v(j)}{(m+1)(p_{k}-1)}%
	\end{equation*}%
	and
	\begin{equation*}
	ESD3_{j}^{U}\left( N,v,P\right) -ESD3_{j}^{U}\left( N,v,P_{-i}\right) =\frac{%
		v(N)-\sum_{j\in N}v(j)}{mp_{k}}-\frac{v(N)-\sum_{j\in N}v(j)}{(m+1)(p_{k}-1)}%
	.
	\end{equation*}%
	
	Finally, the uniqueness is proven in an similar way as the uniqueness in
	Theorem 2 of V\'{a}zquez-Brage et al. (1997).
\end{proof}

\bigskip

\bigskip

Proof of Theorem \ref{esdow}.

\bigskip

\begin{proof}
	Take a TU-game with a priori unions $(N,v,P)\in \mathcal{G}^U$ such that $P=\{P_1,...,P_m\}$ and denote $M=\{1,...,m\}$. First at all, $ESD$ is applied to the modified game (\ref{eq1}), to obtained the reduced game (\ref{eq2});  where note that only the individual pay-offs $u_{r,S}(i)$ and the total pay-off $u_{r,S}(M)$ are necessary. Therefore, for all union $P_{r}$
	
	\begin{align*}
	&w_{r}\left( S\right) =ESD_{r}\left( M,u_{r,S}\right) =u_{r,S}(r)+\frac{%
		u_{r,S}(M)-\sum_{l\in M}u_{r,S}(l)}{m}=\\
	&v\left( S\right) +\frac{v\left( \cup _{k\in M\backslash r}P_{k}\cup S\right)
		-\sum_{l\in M\backslash r}v\left( P_{l}\right) -v\left( S\right) }{m}.
	\end{align*}
	
	Taking again $ESD$, and applying it to the reduced game (\ref{eq2}) as (\ref{eq3}), for all player  $i\in P_{r}$, 
	\begin{align*}
	&ESD_{i}\left( P_{r},w_{r}\right) =w_{r}(i)+\frac{w_{r}(P_{r})-\sum_{t\in
			P_{r}}w_{r}(t)}{p_{r}}=\\
	&v\left( i\right) +\frac{v\left( \cup _{k\in M\backslash r}P_{k}\cup i\right)
		-\sum_{l\in M\backslash r}v\left( P_{l}\right) -v\left( i\right) }{m}+\\
	&\frac{v\left( P_{r}\right) +\frac{v\left( \cup _{k\in M\backslash	r}P_{k}\cup P_{r}\right) -\sum_{l\in M\backslash r}v\left( P_{l}\right)-v\left( P_{r}\right) }{m}-\sum_{t\in P_{r}}(v\left( t\right) +\frac{v\left(\cup _{k\in M\backslash r}P_{k}\cup t\right) -\sum_{l\in M\backslash	r}v\left( P_{l}\right) -v\left( t\right) }{m})}{p_{r}}.\\
	\end{align*}
	
	As
	$$v\left( \cup _{k\in M\backslash	r}P_{k}\cup P_{r}\right) -\sum_{l\in M\backslash r}v\left( P_{l}\right)-v\left( P_{r}\right)= v\left( N\right) -\sum_{l\in M}v\left(	P_{l}\right)$$
	we have that
	\begin{align*}
	&ESD_{i}\left( P_{r},w_{r}\right) =v\left( i\right) +\frac{v\left( \cup _{k\in M\backslash r}P_{k}\cup i\right) }{m}-\frac{v\left( i\right) }{m}-\frac{\sum_{l\in M\backslash r}v\left(	P_{l}\right) }{m}+\frac{v\left( P_{r}\right) }{p_{r}}+\\
	&\frac{v\left( N\right) }{mp_{r}}-\frac{\sum_{l\in M}v\left( P_{l}\right)}{mp_{r}}-\sum_{t\in P_{r}}\frac{v\left(t\right) }{p_{r}}-\sum_{t\in P_{r}}\frac{v\left( \cup _{k\in M\backslash r}P_{k}\cup t\right) }{mp_{r}}+\sum_{t\in P_{r}}\sum_{l\in M\backslash r}\frac{v\left( P_{l}\right) }{mp_{r}}+\sum_{t\in P_{r}}\frac{v\left( t\right) }{mp_{r}}.\\	
	\end{align*}
	
	The second last term can be written as
	\begin{equation*}
	\sum_{t\in P_{r}}\sum_{l\in M\backslash r}\frac{v\left( P_{l}\right) }{mp_{r}%
	}=\sum_{l\in M\backslash r}p_{r}\frac{v\left( P_{l}\right) }{mp_{r}}%
	=\sum_{l\in M\backslash r}\frac{v\left( P_{l}\right) }{m}
	\end{equation*}%
	and then, we obtain that
	\begin{align*}
	&ESD_{i}\left( P_{r},w_{r}\right) =v\left( i\right) +\frac{v\left( \cup _{k\in M\backslash r}P_{k}\cup i\right) 
	}{m}-\frac{v\left( i\right) }{m}+\frac{v\left( P_{r}\right) }{p_{r}}+\\
	&\frac{v\left( N\right) }{mp_{r}}-\frac{%
		\sum_{l\in M}v\left( P_{l}\right) }{mp_{r}}-\sum_{t\in P_{r}}\frac{v\left(
		t\right) }{p_{r}}-\sum_{t\in P_{r}}\frac{v\left( \cup _{k\in M\backslash
			r}P_{k}\cup t\right) }{mp_{r}}+\sum_{t\in P_{r}}\frac{v\left( t\right) }{%
		mp_{r}}.\\
	\end{align*} 
	
	Reordering terms, we have that
	\begin{align*}
	&ESD_{i}\left( P_{r},w_{r}\right) =v\left( i\right) +\frac{1}{m}\left( \sum_{t\in P_{r}}\frac{v\left( t\right) 
	}{p_{r}}-v\left( i\right) \right) +\frac{1}{p_{r}}\frac{v\left( N\right) -\sum_{l\in M}v\left( P_{l}\right) }{m}+\\
	&\frac{1}{p_{r}}\left( v\left( P_{r}\right) -\sum_{t\in P_{r}}v\left(
	t\right) \right) +\frac{1}{m}\left( v\left( \cup _{k\in M\backslash r}P_{k}\cup i\right)
	-\sum_{t\in P_{r}}\frac{v\left( \cup _{k\in M\backslash r}P_{k}\cup t\right) 
	}{p_{r}}\right)\\
	&=ESD4_{i}^{U}\left( N,v,P\right).\\
	\end{align*}
\end{proof}

\bigskip

\bigskip

Proof of Theorem \ref{th10}.

\bigskip

\begin{proof}
	Take a TU-game with a priori unions $(N,v,P)\in \mathcal{G}^U$ such that $P=\{P_1,...,P_m\}$ and denote $M=\{1,...,m\}$. Let us check that $ESD4^{U}$ satisfies BCPA. For all $l\in M$ and all $i,j\in P_l$,
	\begin{align*}
	&ESD4^{U}_{i}\left( N,v,P\right) - ESD4^{U}_{j}\left( N,v,P\right) =ESD2^U_{i}\left( N,v,P\right) + \frac{1}{m}\left( \sum_{t\in P_{l}}\frac{v\left( t\right) 
	}{p_{l}}-v\left( i\right) \right)+ \\
	&\frac{1}{m}\left( v\left( \cup _{k\in M\backslash l}P_{k}\cup i\right)
	-\sum_{t\in P_{l}}\frac{v\left( \cup _{k\in M\backslash l}P_{k}\cup t\right) 
	}{p_{l}}\right)- 
	ESD2^U_{j}\left( N,v,P\right) -\\
	&\frac{1}{m}\left( \sum_{t\in P_{l}}\frac{v\left( t\right) 
	}{p_{l}}-v\left( j\right) \right)- 
	\frac{1}{m}\left( v\left( \cup _{k\in M\backslash l}P_{k}\cup j\right)
	-\sum_{t\in P_{l}}\frac{v\left( \cup _{k\in M\backslash l}P_{k}\cup t\right) 
	}{p_{l}}\right). \\
	\end{align*}%
	
	By the property DMIVIU that satisfies $ESD2^U$ we have that $ESD2_i-ESD2_j=v(i)-v(j)$. Then we have that
	\begin{align*}
	&ESD4^{U}_{i}\left( N,v,P\right) - ESD4^{U}_{j}\left( N,v,P\right) = \\
	&v(i)-v(j)-\frac{v(i)}{m}+\frac{v(j)}{m}+\frac{v\left( \cup _{k\in M\backslash l}P_{k}\cup i\right)}{m}-\frac{v\left( \cup _{k\in M\backslash l}P_{k}\cup j\right)}{m}.
	\end{align*}
	
	On the other hand
	\begin{align*}
	&ESD4^{U}_{i}\left( N\backslash P_{l}\cup i,v_{N\backslash
		P_{l}\cup i},P\backslash P_{l}\cup \{i\}\right) - ESD4^{U}_{j}\left( N\backslash P_{l}\cup j,v_{N\backslash
		P_{l}\cup j},P\backslash P_{l}\cup \{j\}\right)= \\
	&ESD2^U_{i}\left( N\backslash P_{l}\cup i,v_{N\backslash
		P_{l}\cup i},P\backslash P_{l}\cup \{i\}\right) + \frac{1}{m}\left( \frac{v_{N\backslash
			P_{l}\cup i}\left( i\right) 
	}{1}-v_{N\backslash
		P_{l}\cup i}\left( i\right) \right)+ \\
	&\frac{1}{m}\left( v_{N\backslash
		P_{l}\cup i}\left( \cup _{P_k\in P\backslash P_l}P_{k}\cup i\right)
	-\frac{v_{N\backslash
			P_{l}\cup i}\left( \cup _{P_k\in P\backslash P_l}P_{k}\cup i\right) 
	}{1}\right)-\\
	&ESD2^U_{j}\left( N\backslash P_{l}\cup j,v_{N\backslash
		P_{l}\cup j},P\backslash P_{l}\cup \{j\}\right) - \frac{1}{m}\left( \frac{v_{N\backslash
			P_{l}\cup j}\left( j\right) 
	}{1}-v_{N\backslash
		P_{l}\cup j}\left( j\right) \right)- \\
	&\frac{1}{m}\left( v_{N\backslash
		P_{l}\cup j}\left( \cup _{P_k\in P\backslash P_l}P_{k}\cup j\right)
	-\frac{v_{N\backslash
			P_{k}\cup j}\left( \cup _{P_k\in P\backslash P_l}P_{k}\cup j\right) 
	}{1}\right).\\
	\end{align*}
	
	We have that
	\begin{align*}
	&ESD4^{U}_{i}\left( N\backslash P_{l}\cup i,v_{N\backslash
		P_{l}\cup i},P\backslash P_{l}\cup \{i\}\right) - ESD4^{U}_{j}\left( N\backslash P_{l}\cup j,v_{N\backslash
		P_{l}\cup j},P\backslash P_{l}\cup \{j\}\right)=\\
	&ESD2^U_{i}\left( N\backslash P_{l}\cup i,v_{N\backslash
		P_{l}\cup i},P\backslash P_{l}\cup \{i\}\right) - ESD2^U_{j}\left( N\backslash P_{l}\cup j,v_{N\backslash
		P_{l}\cup j},P\backslash P_{l}\cup \{j\}\right)
	\end{align*}
	and then
	
	\begin{align*}
	&ESD2^U_{i}\left( N\backslash P_{l}\cup i,v_{N\backslash
		P_{l}\cup i},P\backslash P_{l}\cup \{i\}\right) - ESD2^U_{j}\left( N\backslash P_{l}\cup j,v_{N\backslash
		P_{l}\cup j},P\backslash P_{l}\cup \{j\}\right)=\\
	&v_{N\backslash
		P_{l}\cup i}(i)+\frac{v_{N\backslash
			P_{l}\cup i}(i)-v_{N\backslash
			P_{l}\cup i}(i)}{1}+
	\frac{v_{N\backslash
			P_{l}\cup i}(N\backslash P_{l}\cup i)-(\sum_{P_k\in P\backslash P_{l}\cup i}v_{N\backslash
			P_{l}\cup i}(P_k))}{m}-\\
	&v_{N\backslash
		P_{l}\cup j}(j)-\frac{v_{N\backslash
			P_{l}\cup j}(j)-v_{N\backslash
			P_{l}\cup j}(j)}{1}-
	\frac{v_{N\backslash
			P_{l}\cup j}(N\backslash P_{l}\cup j)-(\sum_{P_k\in P\backslash P_{l}\cup j}v_{N\backslash
			P_{l}\cup j}(P_k))}{m}=\\
	&v_{N\backslash
		P_{l}\cup i}(i)-v_{N\backslash
		P_{l}\cup j}(j)+\frac{v_{N\backslash
			P_{l}\cup i}(N\backslash P_{l}\cup i)}{m}-\frac{v_{N\backslash
			P_{l}\cup j}(N\backslash P_{l}\cup j)}{m}-\frac{\sum_{P_k\in P\backslash P_{l}\cup i}v_{N\backslash
			P_{l}\cup i}(P_k)}{m}+\\
	&	\frac{\sum_{P_k\in P\backslash P_{l}\cup j}v_{N\backslash
			P_{l}\cup j}(P_k)}{m}=v(i)-v(j)+\frac{v\left( \cup _{k\in M\backslash l}P_{k}\cup i\right)}{m}-\frac{v\left( \cup _{k\in M\backslash l}P_{k}\cup j\right)}{m}-\frac{v(i)}{m}+\frac{v(j)}{m}.
	\end{align*}
	
	Finally, the uniqueness is proven in an similar way as the uniqueness in
	Theorem 2 of V\'azquez-Brage et al. (1997).
\end{proof}

\bigskip

\bigskip

Proof of Theorem \ref{th12}.

\bigskip

\begin{proof}
	Take a TU-game with a priori unions $(N,v,P)\in \mathcal{G}^U$ such that $P=\{P_1,...,P_m\}$ and denote $M=\{1,...,m\}$. Let us check that $ESD5^U$ satisfies QGP. For all $k\in M$, we have that 
	\begin{align*}
	&	\sum_{i\in P_k}ESD5^U_{i}\left( N,v,P\right)=\\
	&\sum_{i\in P_k}\left(\frac{%
		v(P_k)}{p_k}+\frac{v(N)-\sum_{l\in M}v(P_l)}{mp_k}\right)
	+\sum_{i\in P_k}\sum_{\substack{ %
			T\subset P_{k} \\ i\in T}}\frac{P^{m,p_{k},t}}{t}v\left(
	T\right) -\sum_{i\in P_k}\sum_{\substack{ T\subset P_{k} \\ i\notin T}}\frac{P^{m,p_{k},t}}{p_k-t}v\left( T\right)=\\
	&v(P_k)+\frac{v(N)-\sum_{l\in M}v(P_l)}{m}+t\sum_{\substack{ %
			T\subset P_{k}}}\frac{P^{m,p_{k},t}}{t}v\left(
	T\right) 
	-(p_k-t)\sum_{\substack{ T\subset P_{k}}}\frac{P^{m,p_{k},t}}{p_k-t}v\left( T\right)=\\
	&v(P_k)+\frac{v(N)-\sum_{l\in M}v(P_l)}{m}.
	\end{align*}
	
	It is immediate that
	\begin{equation*}
	ESD5^U_{k}\left( M,v/P,P^{m}\right)=v(P_k)+\frac{v(N)-\sum_{l\in M}v(P_l)}{m}.
	\end{equation*}
	
	Let us check that $ESD5^{U}$ satisfies BCU. For all $k\in M$ and all $i,j\in P_k$,
	\begin{align*}
	&ESD5^U_i(N,v,P)-ESD5^U_j(N,v,P)=\\
	&\sum_{\substack{ %
			T\subset P_{k} \\ i\in T}}\frac{P^{m,p_{k},t}}{t}v\left(
	T\right) -\sum_{\substack{ T\subset P_{k} \\ i\notin T}}\frac{P^{m,p_{k},t}}{p_k-t}v\left( T\right)-\sum_{\substack{ %
			T\subset P_{k} \\ j\in T}}\frac{P^{m,p_{k},t}}{t}v\left(
	T\right) +\sum_{\substack{ T\subset P_{k} \\ j\notin T}}\frac{P^{m,p_{k},t}}{p_k-t}v\left( T\right)=\\
	& \sum_{\substack{ %
			T\subseteq P_{k}\backslash j \\ i\in T}}\frac{P^{m,p_{k},t}}{t}v\left(
	T\right)-\sum_{\substack{ %
			T\subseteq P_{k}\backslash i \\ j\in T}}\frac{P^{m,p_{k},t}}{t}v\left(
	T\right)-\sum_{\substack{ T\subset P_{k} \\ i\notin T \\ j\in T}}\frac{P^{m,p_{k},t}}{p_k-t}v\left( T\right) +\sum_{\substack{ T\subset P_{k} \\ j\notin T\\ i\in T}}\frac{P^{m,p_{k},t}}{p_k-t}v\left( T\right)=\\
	&p_k\sum_{\substack{ %
			T\subseteq P_{k}\backslash j \\ i\in T}}\frac{P^{m,p_{k},t}}{(p_k-t)t}v\left(
	T\right)-\sum_{\substack{ %
			T\subseteq P_{k}\backslash i \\ j\in T}}\frac{P^{m,p_{k},t}}{(p_k-t)t}v\left(
	T\right) .
	\end{align*}
	
	On the other hand,
	\small{	\begin{align*}
		&ESD5^U_i\left( N,v,P_{-j}\right)-ESD5^U_j\left( N,v,P_{-i}\right)=\\
		&\frac{v(P_k\backslash j)}{p_k-1}+\frac{v(N)-\sum_{P_l\in P_{-j}}v(P_l)}{(m+1)(p_k-1)}-\frac{v(P_k\backslash i)}{p_k-1}-\frac{v(N)-\sum_{P_l\in P_{-i}}v(P_l)}{(m+1)(p_k-1)}+\sum_{\substack{T\subset P_{k}\backslash j \\ i\in T}}\frac{P^{m+1,p_{k}-1,t}}{t}v\left(
		T\right)\\
		& -\sum_{\substack{ T\subset P_{k}\backslash j \\ i\notin T}}\frac{P^{m+1,p_{k}-1,t}}{p_k-t}v\left( T\right)-\sum_{\substack{ %
				T\subset P_{k}\backslash i \\ j\in T}}\frac{P^{m+1,p_{k}-1,t}}{t}v\left(
		T\right) +\sum_{\substack{ T\subset P_{k}\backslash i \\ j\notin T}}\frac{P^{m+1,p_{k}-1,t}}{p_k-t}v\left( T\right)=\\
		&\frac{v(P_k\backslash j)}{p_k-1}-\frac{v(P_k\backslash i)}{p_k-1}-\frac{v(P_k\backslash j)+v(j)}{(m+1)(p_k-1)}+\frac{v(P_k\backslash i)+v(i)}{(m+1)(p_k-1)}+\sum_{\substack{ %
				T\subset P_{k}\backslash j \\ i\in T}}\frac{P^{m+1,p_{k}-1,t}}{t}v\left(
		T\right)\\
		& -\sum_{\substack{ T\subset P_{k}\backslash j \\ i\notin T}}\frac{P^{m+1,p_{k}-1,t}}{p_k-t}v\left( T\right)-\sum_{\substack{ %
				T\subset P_{k}\backslash i \\ j\in T}}\frac{P^{m+1,p_{k}-1,t}}{t}v\left(
		T\right) +\sum_{\substack{ T\subset P_{k}\backslash i \\ j\notin T}}\frac{P^{m+1,p_{k}-1,t}}{p_k-t}v\left( T\right)=\\
		&\frac{m\cdot v(P_k\backslash j)}{(m+1)(p_k-1)}-\frac{m\cdot v(P_k\backslash i)}{(m+1)(p_k-1)}+\frac{v(i)}{(m+1)(p_k-1)}-\frac{v(j)}{(m+1)(p_k-1)}+\sum_{\substack{ %
				T\subset P_{k}\backslash j \\ i\in T}}\frac{P^{m+1,p_{k}-1,t}}{t}v\left(
		T\right)\\
		& -\sum_{\substack{ T\subset P_{k}\backslash j \\ i\notin T}}\frac{P^{m+1,p_{k}-1,t}}{p_k-t}v\left( T\right)-\sum_{\substack{ %
				T\subset P_{k}\backslash i \\ j\in T}}\frac{P^{m+1,p_{k}-1,t}}{t}v\left(
		T\right)+\sum_{\substack{ T\subset P_{k}\backslash i \\ j\notin T}}\frac{P^{m+1,p_{k}-1,t}}{p_k-t}v\left( T\right)=\\
		&\frac{m\cdot v(P_k\backslash j)}{(m+1)(p_k-1)}-\frac{m\cdot v(P_k\backslash i)}{(m+1)(p_k-1)}+\frac{v(i)}{(m+1)(p_k-1)}-\frac{v(j)}{(m+1)(p_k-1)}+\\
		&\sum_{\substack{ %
				T\subset P_{k}\backslash j \\ i\in T}}\frac{P^{m+1,p_{k}-1,t}}{t}v\left(
		T\right) -\sum_{\substack{ %
				T\subset P_{k}\backslash i \\ j\in T}}\frac{P^{m+1,p_{k}-1,t}}{t}v\left(
		T\right).
		\end{align*}}
	
	Let us see that the two equations are the same. We only need to see that for a player $i$ and any coalition $T\subset P_k\backslash j$ such that $i\in T$, the weights coincide. Consider the all different cases. Noticed that as $i,j\in P_k$, then $p_k\geq 2$.
	
	\textbf{Case i)} $p_k=2$ then
	$$p_k\sum_{\substack{ %
			T\subseteq P_{k}\backslash j \\ i\in T}}\frac{P^{m,p_{k},t}}{(p_k-t)t}v\left(
	T\right)=\frac{p_k}{2(p_k-1)}v(i)=v(i)$$
	and on the other side 
	$$\frac{m\cdot v(P_k\backslash j)}{(m+1)(p_k-1)}+\frac{v(i)}{(m+1)(p_k-1)}=\frac{(m+1)v(i)}{(m+1)(p_k-1)}=v(i).$$
	
	\textbf{Case ii)} $p_k>2$ and $|T|=1$ ($T=i$) then
	$$\frac{p_k}{(p_k-t)t}P^{m,p_{k},t}v\left(i\right)= \frac{p_k}{(p_k-1)}P^{m,p_{k},1}v\left(i\right)=\frac{p_k}{(p_k-1)}\frac{1}{p_k}\left(1+\sum_{j=1}^{p_k-2}\frac{1}{m+j}\right)v(i)$$
	and on the other side 
	$$\frac{v(i)}{(m+1)(p_k-1)}+\frac{P^{m+1,p_{k}-1,t}}{t}v\left(
	i\right)=\frac{v(i)}{(m+1)(p_k-1)}+\frac{1}{p_k-1}\left(1+\sum_{j=1}^{p_k-3}\frac{1}{m+1+j}\right)v(i)$$
	$$= \frac{1}{p_k-1}\left(1+\sum_{j=0}^{p_k-3}\frac{1}{m+1+j}\right)v(i)=\frac{1}{p_k-1}\left(1+\sum_{j=1}^{p_k-2}\frac{1}{m+j}\right)v(i).$$
	
	\textbf{Case ii)} $p_k>2$ and $|T|=p_k-1$ $(T=P_k\backslash j)$ then
	$$p_k\frac{P^{m,p_{k},t}}{(p_k-t)t}v\left(
	P_k\backslash j\right)=\frac{p_k}{p_k-1}\frac{m}{(m+1)p_k}v\left(
	P_k\backslash j\right)=\frac{m}{(m+1)(p_k-1)}v\left(
	P_k\backslash j\right)$$
	and on the other side
	$$ \frac{m\cdot v(P_k\backslash j)}{(m+1)(p_k-1)}.$$
	
	\textbf{Case iii)} $p_k>2$ and $|T|=p_k-2$ then
	$$p_k\frac{P^{m,p_{k},t}}{2(p_k-2)}v\left(
	T\right)=\frac{p_k}{2(p_k-2)}\frac{m+1}{m+2}\frac{1}{p_k-1}\frac{2}{p_k}v\left(
	T\right)=\frac{1}{(p_k-2)}\frac{m+1}{m+2}\frac{1}{p_k-1}v\left(
	T\right)$$
	and on the other side
	$$\frac{P^{m+1,p_{k}-1,t}}{t}v\left(
	T\right)=\frac{1}{p_k-2}\frac{m+1}{m+2}\frac{1}{p_k-1}v\left(
	T\right). $$
	
	\textbf{Case iv)} $p_k>2$ and $|T|=p_k-z$ where $z\in\{3,...,p_k-2\}$
	$$p_k\frac{P^{m,p_{k},t}}{(p_k-t)t}v\left(
	T\right)=\frac{p_k}{z(p_k-z)}\frac{m+(z-1)}{(p_k-(z-1))(m+z)}\left(\sum_{l=0}^{z-2}\frac{p_k-l-t}{p_k-l}\right)v\left(
	T\right)$$
	$$= \frac{1}{(p_k-z)}\frac{m+(z-1)}{(p_k-(z-1))(m+z)}\left(\sum_{l=1}^{z-2}\frac{p_k-l-t}{p_k-l}\right)v\left(
	T\right)$$
	and on the other side
	$$\frac{P^{m+1,p_{k}-1,t}}{t}v\left(
	T\right)=\frac{1}{p_k-z}\frac{m+1+(z'-1)}{(p_k-1-(z'-1))(m+1+z')}\left(\sum_{j=0}^{z'-2}\frac{p_k-1-j-t}{p_k-1-j}\right)$$
	where $z'=z-1$ because $P^{m+1,p_{k}-1,t}$ depends on $P_{k}\backslash j$.
	
	Finally, the uniqueness is proven in an analogous way as the uniqueness in
	Theorem 2 of V\'{a}zquez-Brage et al. (1997).
\end{proof}

\section*{References}

\noindent Alonso-Meijide JM, Fiestras-Janeiro G (2002). Modification of the Banzhaf value for games with a coalition structure. Annals of Operations Research 109, 213-227.\newline
\noindent Alonso-Meijide JM, Costa J, Garc\'{\i}a-Jurado I, Gon\c{c}alves-Dosantos JC (2020). On egalitarian values for cooperative games with a priori unions. TOP 28, 672-688.\newline
\noindent Banzhaf III JF (1964). Weighted voting doesn't work: A mathematical analysis. Rutgers L. Rev., 19, 317.\newline
\noindent B\'eal S, R\'emila E, Solal P (2019). Coalitional desirability and
the equal division value. Theory and Decision 86, 95-106.\newline
\noindent Casajus A. (2009). Outside options, component efficiency, and stability. Games and Economic Behavior 65, 49-61.\newline
\noindent Casajus A, H\"uttner F (2014). Null, nullifying, or dummifying
players: The difference between the Shapley value, the equal division value,
and the equal surplus division value. Economics Letters 122, 167-169.\newline
\noindent Chun Y, Park B (2012). Population solidarity, population
fair-ranking and the egalitarian value. International Journal of Game Theory
41, 255-270.\newline
\noindent Costa J (2016). A polynomial expression of the Owen value in the
maintenance cost game. Optimization 65, 797-809.\newline
\noindent Driessen TSH, Funaki Y (1991). Coincidence of and collinearity 
between game theoretic solutions. OR Spectrum 13, 15-30.\newline
\noindent Ferri\`eres S (2017). Nullified equal loss property and equal
division values. Theory and Decision 83, 385-406.\newline
\noindent G\'omez-R\'ua M, Vidal-Puga J (2010). The axiomatic approach to three values in games with coalition structure. European Journal of Operational Research 207, 795-806.\newline
\noindent Lorenzo-Freire S (2016). On new characterizations of the Owen
value. Operations Research Letters 44, 491-494.\newline
\noindent Owen G (1977) Values of games with a priori unions. In:
Mathematical Economics and Game Theory (R Henn, O Moeschlin, eds.),
Springer, 76-88.\newline
\noindent Owen G (1981) Modification of the Banzhaf-Coleman index for games with a priori unions. In: Power, voting, and voting power. Physica, Heidelberg, 232-238.\newline
\noindent Saavedra-Nieves A, Garc\'{\i}a-Jurado I, Fiestras-Janeiro G
(2018). Estimation of the Owen value based on sampling. In: The Mathematics
of the Uncertain: A Tribute to Pedro Gil (E Gil, E Gil, J Gil, MA Gil,
eds.), Springer, 347-356.\newline
\noindent Shapley LS (1953). A value for n-person games. In: Contributions
to the Theory of Games II (HW Kuhn, AW Tucker, eds.), Princeton University
Press, 307-317.\newline
\noindent Sun P, Hou D, Sun H (2020). The Shapley value for cooperative games with restricted worths. Journal of Mathematical Analysis and Applications, 124762.\newline
\noindent van den Brink R (2007). Null or nullifying players: the difference
between the Shapley value and equal division solutions. Journal of Economic
Theory 136, 767-775.\newline
\noindent van den Brink R, Funaki Y (2009) Axiomatizations of a class of
equal surplus sharing solutions for TU-games. Theory and Decision 67,
303-340.\newline
van den Brink R, Chun Y, Funaki Y, Park B (2016). Consistency, population
solidarity, and egalitarian solutions for TU-games. Theory and Decision 81,
427-447.\newline
\noindent V\'azquez-Brage M, Garc\'ia-Jurado I, Carreras F (1996). The Owen value applied to games with graph-restricted communication. Games and Economic Behavior 12, 45-53.\newline
\noindent V\'{a}zquez-Brage M, van den Nouweland A, Garc\'{\i}a-Jurado I
(1997). Owen's coalitional value and aircraft landing fees. Mathematical
Social Sciences 34, 273-286.\newline
\noindent Winter E (1992). The consistency and potential for values of games with coalition structure. Games and Economic Behavior 4, 132-144.\newline

\end{document}